\documentclass[useAMS,usenatbib]{mn2e}
\usepackage{graphicx}
\usepackage{graphics}
\usepackage{rotating}
\usepackage{amsmath}
\usepackage{amssymb}

\def\USydney{$^{1}$}
\def\CAASTRO{$^{2}$}
\def\CASS{$^{3}$}
\def\UWisc{$^{4}$}
\def\ANU{$^{5}$}
\def\Curtin{$^{6}$}
\def\CfA{$^{7}$}
\def\SKASA{$^{8}$}
\def\Rhodes{$^9$}
\def\ASU{$^{10}$}
\def\Haystack{$^{11}$}
\def\RRI{$^{12}$}
\def\MIT{$^{13}$}
\def\UW{$^{14}$}
\def\Victoria{$^{15}$}
\def\UMichigan{$^{16}$}
\def\Tata{$^{17}$}
\def\NRAO{$^{18}$}
\def\UMelbourne{$^{19}$}

\title[Limits on low frequency radio emission from southern exoplanets]{Limits on low frequency radio emission from southern exoplanets with the Murchison Widefield Array}
\author[Murphy et al. ]{\parbox[t]{\textwidth}
{Tara Murphy\USydney$^,$\CAASTRO\thanks{E-mail: tara.murphy@sydney.edu.au}, 
Martin E. Bell\USydney$^,$\CAASTRO$^,$\CASS, David L. Kaplan\UWisc, B. M. Gaensler\USydney$^,$\CAASTRO, Andr{\'e} R. Offringa\ANU$^,$\CAASTRO, Emil Lenc\USydney$^,$\CAASTRO, Natasha Hurley-Walker\Curtin,
G.~Bernardi\CfA$^,$\SKASA$^,$\Rhodes,
J.~D.~Bowman\ASU, 
F.~Briggs\ANU$^,$\CAASTRO, 
R.~J.~Cappallo\Haystack, 
B.~E.~Corey\Haystack, 
A.~A.~Deshpande\RRI, 
D.~Emrich\Curtin,
R.~Goeke\MIT,
L.~J.~Greenhill\CfA,
B.~J.~Hazelton\UW, 
J.~N.~Hewitt\MIT, 
M.~Johnston-Hollitt\Victoria,
J.~C.~Kasper\UMichigan$^,$\CfA, 
E.~Kratzenberg\Haystack, 
C.~J.~Lonsdale\Haystack, 
M.~J.~Lynch\Curtin, 
S.~R.~McWhirter\Haystack,
D.~A.~Mitchell\CASS$^,$\CAASTRO, 
M.~F.~Morales\UW, 
E.~Morgan\MIT, 
D.~Oberoi\Tata, 
S.~M.~Ord\Curtin$^,$\CAASTRO,
T.~Prabu\RRI, 
A.~E.~E.~Rogers\Haystack, 
D.~A.~Roshi\NRAO 
N.~Udaya~Shankar\RRI, 
K.~S.~Srivani\RRI, 
R.~Subrahmanyan\RRI$^,$\CAASTRO, 
S.~J.~Tingay\Curtin$^,$\CAASTRO, 
M.~Waterson\Curtin$^,$\ANU,
R.~B.~Wayth\Curtin$^,$\CAASTRO, 
R.~L.~Webster\UMelbourne$^,$\CAASTRO, 
A.~R.~Whitney\Haystack, 
A.~Williams\Curtin, 
C.~L.~Williams\MIT}\\
\vspace*{1pt} \\
$^{1}$Sydney Institute for Astronomy, School of Physics, The University of Sydney, NSW 2006, Australia\\
$^{2}$ARC Centre of Excellence for All-sky Astrophysics (CAASTRO)\\
$^{3}$CSIRO Astronomy and Space Science, Marsfield, NSW 2122, Australia\\
$^{4}$Department of Physics, University of Wisconsin--Milwaukee, Milwaukee, WI 53201, USA\\
$^{5}$Research School of Astronomy and Astrophysics, Australian National University, Canberra, ACT 2611, Australia\\
$^{6}$International Centre for Radio Astronomy Research, Curtin University, Bentley, WA 6102, Australia\\
$^{7}$Harvard-Smithsonian Center for Astrophysics, Cambridge, MA 02138, USA\\
$^{8}$SKA SA, 3rd Floor, The Park, Park Road, Pinelands, 7405, Cape Town, South Africa \\
$^{9}$Department of Physics and Electronics, Rhodes University, Grahamstown, 6140, South Africa \\
$^{10}$School of Earth and Space Exploration, Arizona State University, Tempe, AZ 85287, USA\\
$^{11}$MIT Haystack Observatory, Westford, MA 01886, USA\\
$^{12}$Raman Research Institute, Bangalore 560080, India\\
$^{13}$Kavli Institute for Astrophysics and Space Research, Massachusetts Institute of Technology, Cambridge, MA 02139, USA\\
$^{14}$Department of Physics, University of Washington, Seattle, WA 98195, USA\\
$^{15}$School of Chemical \& Physical Sciences, Victoria University of Wellington, Wellington 6140, New Zealand \\
$^{16}$Department of Atmospheric, Oceanic and Space Sciences, University of Michigan, Ann Arbor, MI 48109, USA\\
$^{17}$National Centre for Radio Astrophysics, Tata Institute for Fundamental Research, Pune 411007, India\\
$^{18}$National Radio Astronomy Observatory, Charlottesville and Greenbank, USA\\
$^{19}$School of Physics, The University of Melbourne, Parkville, VIC 3010, Australia\\
}

\begin{document}

\date{January 2014}

\pagerange{\pageref{firstpage}--\pageref{lastpage}} \pubyear{2014}

\maketitle

\label{firstpage}

\begin{abstract}
We present the results of a survey for low frequency radio emission from 17 known exoplanetary systems with the 
Murchison Widefield Array.
This sample includes 13 systems that have not previously been targeted with radio observations.
We detected no radio emission at 154~MHz, and put $3\sigma$ upper limits in the range 15.2--112.5 mJy on this emission. 
We also searched for circularly polarised emission and made no detections, obtaining $3\sigma$ upper limits in the range 3.4--49.9 mJy.
These are comparable with the best low frequency radio limits in the existing literature
and translate to luminosity limits of between $1.2\times10^{14}$~W and 
$1.4\times10^{17}$~W if the emission is assumed to be $100\%$ circularly polarised. These are the first results from a
larger program to systematically search for exoplanetary emission with the MWA. 
\end{abstract}


\section{Introduction}
Magnetised extrasolar planets are expected to emit strongly at radio wavelengths, in
the same way as magnetised planets in our own solar system \citep{winglee86,zarka01}. Intense emission
can be generated by the electron-cyclotron maser instability if the planet has an intrinsic
magnetic field and a source of energetic electrons \citep{dulk85}. This emission is sporadic, with variability 
timescales spanning seconds to days. Unlike in the optical regime, in which
radiation from planets is much weaker than that of their parent star, in radio it can be of comparable
strength: Jupiter's radio emission in the decametre
band is as intense as solar radio bursts \citep{zarka01}.
The strongest exoplanetary emission is likely to be from planets
that are more massive than Jupiter, orbiting their parent star at short orbital 
distances.

A majority of exoplanets have been discovered through indirect means such as radial velocity and
transit searches \citep{perryman11}.
A small number of known exoplanets\footnote{See http://exoplanet.eu/catalog for a complete catalogue} have 
been detected through direct imaging, and radio observations provide another method of making direct detections.
Radio observations of exoplanets would allow us to confirm that the planet has a magnetic field, and put
a limit on the magnetic field strength near the surface of the planet \citep{hess11}.
The detection of circular polarisation would indicate which magnetic hemisphere the emission comes from 
and would give a limit on the plasma density in the magnetosphere \citep{bastian00}.

Early attempts to detect radio emission from exoplanets occurred before the discovery of 
the first exoplanet around a main sequence star, 51~Peg, by \citet{mayor95}. For example, \citet{winglee86} targeted six nearby stars 
with the Very Large Array (VLA) at 333 and 1400~MHz, but made no detections.
More recently, there have been a number of attempts to detect radio emission from known exoplanets.
\citet{bastian00} conducted a search of seven exoplanets with the VLA at 333 and 1465~MHz and one at 74~MHz. 
They made no detections, with typical $3\sigma$ upper limits of 0.06--0.21~mJy at 1465~MHz, 3--30~mJy at 
333~MHz and 150~mJy at 74~MHz. \citet{lazio04} observed five exoplanets with the VLA at 74~MHz, as part of
the VLA Low Frequency Sky Survey (VLSS), finding $3\sigma$ upper limits of 262--390~mJy.
\citet{george07} targeted $\epsilon$ Eri and HD~128311 at 150~MHz with the Giant Metrewave Radio Telescope (GMRT). 
They reported $3\sigma$ limits of 9.4~mJy and 18.6~mJy respectively for the two systems.
\citet{smith09} targeted HD~189733 during the planet's secondary eclipse, reporting a $3\sigma$ upper limit of 81~mJy
in the frequency range 307--347~MHz.
\citet{lazio10} targeted HD~80606b with the VLA and reported $3\sigma$ limits of 1.7~mJy at 
325~MHz and $48 \mu$Jy at 1425~MHz.
\citet{stroe12} observed $\tau$ Bootis at 1.7~GHz with the Westerbork Synthesis Radio Telescope, and reported
a $3\sigma$ upper limit of 0.13~mJy on emission from the exoplanetary system. \citet{hallinan13} observed $\tau$ Bootis with the GMRT, reporting a $3\sigma$ upper
limit of 1.2~mJy at 150~MHz.

GMRT observations of the Neptune-mass exoplanet HAT-P-11b by \citet{lecavelier13} show a possible detection of
radio emission. If the emission is associated with the planet, then the flux density is 3.87~mJy at 150~MHz. However, they reported that the detection was not confirmed in follow-up observations, and deeper observations are required.
The largest radio survey of exoplanets to date was conducted by \citet{sirothia14}, with data from the TIFR GMRT 
Sky Survey. No detections were made, with 150~MHz $3\sigma$ upper limits of between 8.7 and 136~mJy placed on 171 planetary systems.
In summary, there have been no confirmed detections of planetary systems at radio wavelengths to date.

Cyclotron maser emission has a maximum frequency determined by the electron gyrofrequency and is 
proportional to the magnetic field strength \citep[see Equation~\ref{eqf};][]{farrell99}.
Cyclotron maser emission is $100\%$ circularly polarised, and so planets should 
be detectable in circular polarisation (Stokes V) at similar levels to their total intensity (Stokes I) emission.
Exoplanetary radio emission is expected to peak at frequencies less than $10-100$~MHz \citep[see, for example,][]{griessmeier07} and hence
has been inaccessible by most telescopes. 

In this paper we present a search for low frequency radio emission from known exoplanets with the Murchison Widefield Array \citep[MWA;][]{lonsdale09,tingay13,bowman13}. We have targeted 17 exoplanetary systems that fall within the region of the MWA Transients Survey (MWATS), a blind survey 
for transients and variables. The MWA sensitivity is confusion limited in Stokes I. 
However, we take advantage of the low density of circularly polarised sources and hence the improved sensitivity in 
Stokes V to conduct deeper searches for polarized radio emission from these sources.
Using the assumption that the circularly polarised
emission is a large fraction of the total intensity, these limits can be used to calculate luminosity limits. 
Note that this also ignores the likely time variability of exoplanetary radio emission.

\section{Observations and Data Analysis}\label{s_obs}
The Murchison Widefield Array is a 128-tile low-frequency radio interferometer located in Western Australia. It operates between 80 and 300~MHz with a processed bandwidth of 30.72 MHz for both linear polarisations. The 128 tiles are distributed over a $\sim$ 3-km diameter area, with a minimum baseline of 7.7~m and a maximum baseline of 2864~m. MWA operations began in 2013 June. See \citet{hurleywalker14} and \citet{bell14} for a description of the MWA observing modes.

We obtained observations between 2013-07-09 and 2014-06-13 (UTC) as part of the MWATS survey (Bell et al., in prep).
Three different meridian declination strips at Dec $= 1.6^{\circ}$, $-26.7^{\circ}$ 
(zenith) and $-55^{\circ}$ were observed in drift scan mode for a whole night, at a cadence of approximately once 
per month 
(at 154~MHz). Each declination was observed in turn with a integration time of two minutes. The $\sim25^\circ$ field of view (equivalent to about 2~hours of sidereal time) means that each observation overlaps with the previous one at the same declination. The specific declination 
strips that contained our target sources for this work are summarised in Table~\ref{obs_table}.

\subsection{Data reduction}
Our data reduction and analysis procedure follows the approach described by \citet{bell14}. 
Each declination strip was calibrated using a two minute observation of a bright, well modeled source. A single 
time-independent, frequency dependent amplitude and phase calibration solution was applied to all observations for 
a given night (per declination strip). 
For each snapshot observation, the visibilities were preprocessed with the \textsc{Cotter} MWA preprocessing pipeline, which 
flags radio-frequency interference using the \textsc{AOFlagger} \citep{offringa12}, averages the data 
and converts the data to the CASA measurement set format.
The observations were then imaged and cleaned 
using the WSClean algorithm \citep{offringa14}. A pixel size of $0\farcm75$ and image size of 3072 pixels was used for imaging.

The WSClean algorithm was used to produce both Stokes I images with Briggs weighting $-1$ (closer to uniform weighting) 
and Stokes V images with Briggs weighting $+1$ (closer to natural weighting). 
The WSClean algorithm does this by forming a $2\times2$ complex Jones matrix $I$ for each image pixel. Beam correction is then applied 
by inverting the beam voltage matrix $B$, and computing $B^{-1} I B^{*-1}$ where $*$ denotes the conjugate transpose. A full description of this approach is given
by \citet{offringa14}. Note that some aspects of the MWA data reduction process are 
being improved, and so future work is likely to have somewhat better sensitivity than presented here.

For each of the declination strips the resulting images (in Stokes I and Stokes V) were mosaiced together (co-added and weighted by the primary beam) to 
increase the sensitivity, as described by \citet{hurleywalker14}. Our final data product consisted of a series of snapshot images with two minute integrations, 
and a mosaiced image that combined all the snapshot images for a single night of observing (the number of snapshots in each
night is given in Table~\ref{obs_table}).

\begin{table}
\centering
\caption{Observations used in this paper. The name of each epoch consists of the month of observation and the declination strip observed. N gives the number of individual snapshot observations taken that night.}
\begin{tabular}{|l|c|c|c|}
\hline
Epoch & Start Date (UTC) & End Date (UTC) & N  \\
\hline
Sep $-55^\circ$ & 2013-09-16 13:30:39 & 2013-09-16 21:24:39 & 77 \\
Dec $-55^\circ$ & 2013-12-06 13:53:27 & 2013-12-06 20:05:27 & 60 \\
Dec $+1.6^\circ$ & 2013-12-06 13:51:27 & 2013-12-06 20:09:27 & 61 \\
Apr $+1.6^\circ$ & 2014-04-28 10:51:59 & 2014-04-28 20:45:59 & 96 \\
Apr $-26^\circ$ & 2014-04-28 10:49:59 & 2014-04-28 20:43:51 & 96 \\
Apr $-55^\circ$ & 2014-04-28 10:47:59 & 2014-04-28 20:41:59 & 96 \\
Jun $-26^\circ$ & 2014-06-09 11:16:15 & 2014-06-09 20:56:15 & 55 \\
Jun $-55^\circ$ & 2014-06-12 11:04:31 & 2014-06-13 20:20:23 & 74 \\
\hline
\label{obs_table}
\end{tabular}
\end{table}

\subsection{Sample Selection}
Of the 1110 confirmed exoplanetary systems\footnote{From the exoplanet.eu catalogue, as of 2014 May 14.}, 
347 fall within the region covered by the MWATS survey as at 2014 June.
We calculated the expected maximum emission frequency and flux density using the models of \citet{lazio04} and 
selected the sources for which these parameters were close to or above the MWA detection capabilities.
We also included the sources listed by \citet{nichols12} as the ten most likely candidates for radio emission generated 
by magnetosphere-ionosphere coupling. This resulted in a sample of 17 of the most likely candidates for detectable 
emission at MWA frequencies. Fifteen of these sources are in the southern hemisphere and, as most previous studies have
focused on northern samples, have not previously been targeted with radio observations.
Table~\ref{t_sample} lists the key properties of the exoplanetary systems in our sample. 
\begin{table*}
\caption{Properties of exoplanetary systems targeted in our survey. $N_p$ is the number of known planets in the system. Where there
are multiple planets, the properties are given for the first discovered planet (which in all cases except Gliese 876~b is the planet closest to its host star). {\it a} is the semi-major axis and $D$ is the distance to the star. S$_{150}$ is best low frequency (150~MHz) $3\sigma$ flux
limit from the literature where available \citep[Geo07:][]{george07} and \citep[Sir14:][]{sirothia14}. Note that \citet{sirothia14} report a 120~mJy source at $3\farcs2$ from the position of 61 Vir. However, it is likely this is a background source unassociated with the exoplanetary
system (see text for more details).}
\label{t_sample}
\begin{tabular}{lcccrcrrrrr}
\hline  
Name & RA & Dec               & Discovery & D    & $N_p$ & Mass    & Period & a & S$_{150}$ & Reference \\
     & (J2000)   & (J2000)    & Year      & (pc) &       & ($M_J$) & (days) & (AU) & (mJy) &  \\
 \hline
WASP-18~b & 01:37:25 & $-$45:40:40 & 2009 & 100.0 & 1 & 10.43 & 0.9 & 0.02 & & \\
Gl 86~b & 02:10:25 & $-$50:49:25 & 2000 & 10.9 & 1 & 4.01 & 15.8 & 0.11 & & \\
$\epsilon$ Eri~b & 03:32:55 & $-$09:27:29 & 2000 & 3.2 & 2 & 1.55 & 2502.0 & 3.39 & $<$9.36 & Geo07 \\
HD 37605~b & 05:40:01 & $+$06:03:38 & 2004 & 44.0 & 2 & 2.81 & 55.0 & 0.28 & & \\
HD 41004 B~b & 05:59:49 & $-$48:14:22 & 2004 & 43.0 & 1 & 18.40 & 1.3 & 0.02 & & \\
HD 102365~b & 11:46:31 & $-$40:30:01 & 2011 & 9.2 & 1 & 0.05 & 122.1 & 0.46 & & \\
61 Vir~b & 13:18:23 & $-$18:18:39 & 2009 & 8.5 & 3 & 0.02 & 4.2 & 0.05 & & \\
HD 128311~b & 14:36:00 & $+$09:44:47 & 2002 & 16.6 & 2 & 2.18 & 448.6 & 1.10 & $<$18.60 & Geo07 \\
HIP 79431~b & 16:12:41 & $-$18:52:32 & 2010 & 14.4 & 1 & 2.10 & 111.7 & 0.36 & $<$11.40 & Sir14 \\
HD 147018~b & 16:23:00 & $-$61:41:20 & 2009 & 43.0 & 2 & 2.12 & 44.2 & 0.24 & & \\
HD 147513~b & 16:24:01 & $-$39:11:34 & 2003 & 12.9 & 1 & 1.21 & 528.4 & 1.32 & & \\
GJ 674~b & 17:28:40 & $-$46:53:43 & 2007 & 4.5 & 1 & 0.04 & 4.7 & 0.04 & & \\
HD 162020~b & 17:50:37 & $-$40:19:05 & 2002 & 31.3 & 1 & 14.40 & 8.4 & 0.07 & & \\
HD 168443~b & 18:20:04 & $-$09:35:34 & 1998 & 37.4 & 2 & 7.66 & 58.1 & 0.29 & & \\
Gl 785~b & 20:15:16 & $-$27:01:59 & 2010 & 8.9 & 1 & 0.05 & 74.7 & 0.32 & $<$14.70 & Sir14 \\
GJ 832~b & 21:33:34 & $-$49:00:32 & 2008 & 4.9 & 1 & 0.64 & 3416.0 & 3.40 & & \\
Gliese 876~b & 22:53:13 & $-$14:15:12 & 2000 & 4.7 & 4 & 1.93 & 61.0 & 0.21 & & \\

\hline

\end{tabular}
\end{table*}

\section{Results and Discussion}\label{s_results}
We searched the individual snapshot images and mosaiced images in Stokes I and Stokes V for all of the exoplanetary
systems in our sample. No radio sources were detected at the positions of any of the systems considered, within the
MWA position errors.
For each source we measured the upper limit on the Stokes I and Stokes V emission in
each snapshot image and in the mosaiced image by calculating the rms in a box several times the synthesised beam, centred on the source position. 
Table~\ref{t_results} shows the best measured limits from the 
mosaiced images (effectively limits on steady emission from these systems). Each target source typically appeared in $\sim 20-30$ snapshot images, so the limits in the snapshot
images (which have an integration time of 2 minutes and a cadence of 6 minutes) were typically a factor of $\sim 4-6$ times higher than the mosaiced images.

Note that an imperfect MWA primary beam model means that flux density measurements far from the phase centre of an observation could have errors of up to $10\%$ \citep{hurleywalker14}. 
The polarisation leakage has been estimated empirically by measuring the polarisation of bright sources as they pass 
through the primary beam. The leakage is typically $1-2\%$ at zenith
and up to $3-5\%$ at low elevations where the beam is less well modelled (Sutinjo et al. submitted).
\begin{table*}
\caption{Our measured $3\sigma$ flux density $S_{154}$ upper limits and derived luminosity $L$ limits for Stokes I total intensity emission and Stokes V circularly polarised emission, and comparison to theoretical predictions from the literature.}\label{t_results}
\begin{tabular}{lrrrrrrrrrr}
\hline
Name & \multicolumn{4}{c}{MWA 154 MHz} &  \multicolumn{2}{c}{Lazio et al. 2004} & \multicolumn{2}{c}{Grie{\ss}meier et al. 2011} & \multicolumn{2}{c}{Nichols 2012} \\
     & $S_{154}$ (I) & L (I) & $S_{154}$ (V)  & L (V) & $\nu_c$  & $S_\nu$  & $\nu_c$ & $S_\nu$ &  $S_{1,24}$ & $S_{3,54}$   \\
 & (mJy) & (W) & (mJy) & (W) & (MHz) & (mJy) & (MHz) & (mJy) & (mJy) & (mJy) \\
\hline
WASP-18  & $<15.2$ & $<2.2\times 10^{17}$  & $<3.4$ & $<5.1\times 10^{16}$  & 812 & 6.8 & 92 & 50 & --- & --- \\
Gl 86  & $<16.4$ & $<2.9\times 10^{15}$  & $<4.6$ & $<8.1\times 10^{14}$  & 464 & 36.7 & 61 & 4 & 0.05 & 1.57 \\
$\epsilon$ Eridani  & $<51.0$ & $<7.7\times 10^{14}$  & $<7.7$ & $<1.2\times 10^{14}$  & 95 & 2.4 & 33 & 20 & 0.61 & 18.21 \\
HD 37605  & $<93.0$ & $<2.6\times 10^{17}$  & $<10.9$ & $<3.1\times 10^{16}$  & 257 & 0.6 & 49 & 2 & --- & --- \\
HD 41004 B  & $<68.6$ & $<1.9\times 10^{17}$  & $<14.5$ & $<4.0\times 10^{16}$  & 1832 & 33.5 & 172 & 600 & --- & --- \\
HD 102365  & $<45.3$ & $<5.7\times 10^{15}$  & $<9.0$ & $<1.1\times 10^{15}$  & $<1$ & 22.7 & 3 & 0 & 0.07 & 2.17 \\
61 Vir  & --- & ---  & $<6.6$ & $<7.0\times 10^{14}$  & $<1.0$ & 2206.0 & --- & --- & 0.08 & 2.55 \\
HD 128311  & $<43.7$ & $<1.8\times 10^{16}$  & $<10.0$ & $<4.0\times 10^{15}$  & 168 & 0.5 & 41 & 1 & --- & --- \\
HIP 79431  & $<50.0$ & $<1.5\times 10^{16}$  & $<13.7$ & $<4.2\times 10^{15}$  & 158 & 3.9 & 40 & 0 & --- & --- \\
HD 147018  & $<76.3$ & $<2.1\times 10^{17}$  & $<49.9$ & $<1.4\times 10^{17}$  & 160 & 0.8 & 40 & 0 & --- & --- \\
HD 147513  & $<73.1$ & $<1.8\times 10^{16}$  & $<24.2$ & $<5.9\times 10^{15}$  & 63 & 0.7 & 27 & 2 & --- & --- \\
GJ 674  & $<42.4$ & $<1.3\times 10^{15}$  & $<14.6$ & $<4.4\times 10^{14}$  & $<1$ & 8975.5 & 1 & 5000 & 0.30 & 9.04 \\
HD 162020  & $<85.6$ & $<1.2\times 10^{17}$  & $<27.4$ & $<3.9\times 10^{16}$  & 191 & 10.3 & 145 & 8 & --- & --- \\
HD 168443  & $<112.5$ & $<2.3\times 10^{17}$  & $<33.2$ & $<6.8\times 10^{16}$  & 1365 & 0.5 & 97 & 0 & --- & --- \\
Gl 785  & $<38.7$ & $<4.5\times 10^{15}$  & $<11.2$ & $<1.3\times 10^{15}$  & $<1$ & 42.7 & 4 & 0 & 0.08 & 2.39 \\
GJ 832  & $<16.8$ & $<6.0\times 10^{14}$  & $<4.7$ & $<1.7\times 10^{14}$  & 21 & 1.4 & 18 & 0 & 0.25 & 7.59 \\
Gliese 876  & $<66.3$ & $<2.1\times 10^{15}$  & $<17.3$ & $<5.6\times 10^{14}$  & 136 & 91.0 & 42 & 6 & 0.28 & 8.38 \\

\hline
\end{tabular}
\end{table*}

\begin{figure*}
\includegraphics[width=14cm]{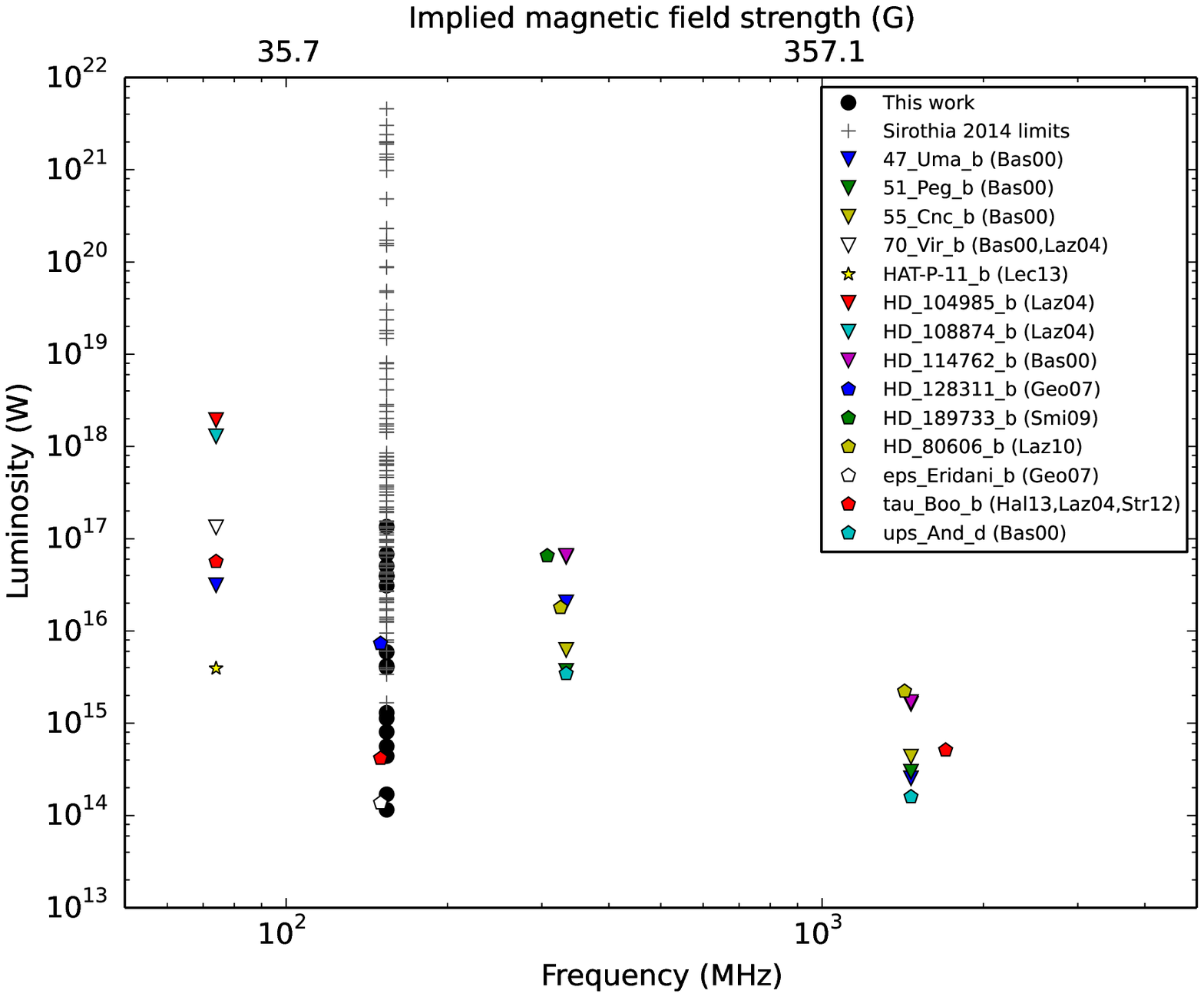}
\caption{Luminosity limits on radio emission from exoplanets. The limits from this work, calculated from our Stokes V measurements,
are shown as black circles. The references in the legend are: Bas00 \citep{bastian00}; Laz04 \citep{lazio04}; Geo07 \citep{george07}; Smi09 \citep{smith09}; Laz10 \citep{lazio10}; Str12 \citep{stroe12}; Hal13 \citep{hallinan13}; Lec13 \citep{lecavelier13}. The secondary x-axis gives the implied magnetic field strength $B_p^{max}$ calculated using Equation (1).}\label{f_results}
\end{figure*}

We calculated luminosity limits from steady emission, assuming an emission bandwidth equal to the observing frequency (154~MHz) and 
a solid angle of $4\pi$~sr (i.e. we are ignoring any beaming of the emission).
These are shown in Columns 3 and 5 of Table~\ref{t_results} for Stokes I and Stokes V respectively. 
For comparison, we have also converted the limits published in the literature using the same method. These are
shown in Figure~\ref{f_results}, with our new limits as black circles.

The maximum emission frequency of cyclotron maser emission  is at the electron gyrofrequency and is 
proportional to the maximum planetary magnetic field strength, $B_p^{max}$ \citep{farrell99}:
\begin{equation}\label{eqf}
f_c^{max} = \frac{eB_p^{max}}{2\pi m_e} = 2.8 B_p^{max} .
\end{equation}
Hence a detection at 154~MHz would imply a magnetic field strength of $B_p = 55$ G.

The Jovian magnetic field, predicted by models based on spacecraft observations of Jupiter,
ranges in strength from 2 to 14~G \citep{connerney93}. Hence, for an exoplanet to be detected
at 154~MHz, it would need a magnetic field strength approximately four times that of Jupiter's.
Models of convection-driven dynamos in planets predict that young,
giant extrasolar planets of 5--10 Jupiter masses could have a surface magnetic field 
strength 5--12 times larger than Jupiter's surface magnetic field \citep{christensen06,christensen09}.

We have measured $3\sigma$ flux density limits between 3.4 and 49.9~mJy in Stokes V (see Table~\ref{t_results}). 
The high values for several of the Stokes V limits are due to contamination from a nearby bright source or Galactic plane emission. 
Assuming $100\%$ circularly
polarised emission, these translate to luminosity limits between $5.8\times10^{13}$~W and $6.8\times10^{16}$. 
These are comparable to the best radio limits given in the literature for
$\epsilon$~Eri \citep{george07} and 47~UMa \citep{bastian00}.
Jupiter generates between
$10^{10}$ and $10^{11}$ W of power between 1 and 40~MHz \citep{zarka04}, making our best limits about 3 orders of
magnitude greater.

Table~\ref{t_results} also lists predicted radio flux densities from various models presented in the literature. Columns 6 and 7 give 
predictions for the characteristic emission frequency $\nu_c$ (in MHz) and the burst flux density at that frequency $S_\nu$ (in mJy) from \citet{lazio04}.
These are calculated from the radiometric Bode's law and Blackett's law \citep[see][]{lazio04}, and then applying the assumption that burst flux density
could be factor of 100 greater than these calculations. Our measured limits for Gl~86, GJ~674, 61~Vir, HD~41004~B, Gl~785 and Gliese~876 are lower than these predictions, although note that the characteristic emission
frequency is predicted to be outside the MWA frequency range for all but two of these.

\citet{griessmeier07,griessmeier11} present predictions based on three models of exoplanetary emission: a magnetic energy model, a kinetic energy model, and a model in which exoplanets are assumed to be subject to frequent stellar eruptions similar to solar coronal mass ejections. Columns 8 and 9 of Table~\ref{t_results} gives
the maximum predicted emission frequency $\nu$ (in MHz) and the corresponding flux density $S_\nu$ (in mJy) from these three models. 
For the three sources that have a predicted emission frequency within the range of the MWA (WASP-18, HD~41004~B and HD~162020) we provide constraining limits for the first two.

Finally, \citet{nichols12} presents predictions based on a model of magnetosphere-ionosphere coupling in Jupiter-like
exoplanets. \citet{nichols12} gives predictions for the maximum flux density based on different values of dynamic
pressure and planetary angular velocity. Column 10 gives the predicted flux density based on solar dynamic pressure and angular
velocity ($S_1$, at 24~MHz, in mJy). Column 11 gives the predicted flux density based on solar dynamic pressure and $3\times$-solar angular
velocity ($S_3$, at 54~MHz, in mJy). These predictions are all below our observed limits.

One of the sources in our target list, 61~Vir, was discussed by \citet{sirothia14}, who reported a 120~mJy source at 150~MHz, located $3\farcs2$ from the position of the star. We detect a source at 187~mJy at 
the same position. This is an NVSS source (NVSS J131823-181851) with a flux density of 50.9~mJy at 1.4~GHz \citep{condon98} and is also detected in the Westerbork In the Southern Hemisphere (WISH) survey with a flux density of 159~mJy at 325~MHz \citep{debreuck02}. The position of the source in NVSS, WISH and our MWA observations agree to within the respective
survey errors. The high proper motion of 61~Vir \citep[($1070.36$, $-1063.69$) mas/year;][]{vanleeuwen07} argues against this radio source being associated with the star or planetary system. When the NVSS images were taken (between 1993 and 1996), the proper motion corrected position of 61~Vir would have been offset by at least $22\farcs3$ from the NVSS position, well outside the NVSS positional error of $\sim 1^{\prime\prime}$. Hence
we conclude that this is a background radio source that is not associated with 61~Vir.

\section{Conclusion}\label{s_conclusion}
We have presented the first results from a survey for low frequency radio emission from 17 known exoplanets with the 
Murchison Widefield Array.
This sample includes 13 exoplanets that have not previously been targeted with radio observations.
We made no detections of radio emission at 154~MHz, and put upper limits in the range 15.2--112.5 mJy on this 
emission. 
We also searched for circularly polarised emission and made no detections, putting upper limits in the range 3.4--49.9 mJy.
These are comparable with the best low frequency radio limits in the existing literature and translate to luminosity 
limits in the range of $5.8\times10^{13}$~W to
$6.8\times10^{16}$~W if the emission is assumed to be $100\%$ circularly polarised.

As discussed by \citet{bastian00}, there are a number of reasons which may explain why we have not made any detections.
The most obvious is that we need more sensitive observations, as only our best limits place any constraints on predicted
flux densities. The second issue is that our observing frequency, while lower than many previous observations of exoplanetary 
systems, is still too high compared to the predicted maximum emission frequency for many systems in our sample. These
instrumental limitations will be reduced with future telescopes, and ultimately the Square Kilometre Array low frequency instrument \citep{lazio09}.

In addition to these limitations, we need observations with better coverage of the planetary orbital period. 
Radio emission
from exoplanetary systems is likely to be orbitally beamed, and so full coverage of a planetary orbital period would put
a more stringent limit on the emission. Since we do not know the axis of beaming, it is critical to target a range of
known exoplanets. As well as being modulated by the orbital frequency, it is likely that the emission is time variable. Hence the ideal observing program would involve constant monitoring of a large number of exoplanetary systems. 

In future work with the MWA we will use lower frequency (90~MHz) observations and cover the full orbital period of several known
systems to increase the probability of detection.
There are also projects underway with LOFAR to search for exoplanetary radio emission \citep{zarka09}.
The development of the Square Kilometre Array low frequency instrument (SKA1-Low)
will provide an opportunity to detect exoplanets at low frequencies (a Jupiter-like planet could be detected out to
$\sim10$~pc), and enable blind surveys which have
the potential to discover new exoplanetary systems through their radio emission \citep{zarka14}.

\section*{Acknowledgments}
We thank Elaine Sadler for useful discussions and Aina Musaeva for assistance with the data reduction.
This scientific work makes use of the Murchison Radio-astronomy Observatory, operated by CSIRO. We acknowledge the Wajarri Yamatji people as the traditional owners of the Observatory site. Support for the MWA comes from the U.S. National Science Foundation (grants AST-0457585, PHY-0835713, CAREER-0847753, and AST-0908884), the Australian Research Council (LIEF grants LE0775621 and LE0882938), the U.S. Air Force Office of Scientific Research (grant FA9550-0510247), and the Centre for All-sky Astrophysics (an Australian Research Council Centre of Excellence funded by grant CE110001020). Support is also provided by the Smithsonian Astrophysical Observatory, the MIT School of Science, the Raman Research Institute, the Australian National University, and the Victoria University of Wellington (via grant MED-E1799 from the New Zealand Ministry of Economic Development and an IBM Shared University Research Grant). The Australian Federal government provides additional support via the Commonwealth Scientific and Industrial Research Organisation (CSIRO), National Collaborative Research Infrastructure Strategy, Education Investment Fund, and the Australia India Strategic Research Fund, and Astronomy Australia Limited, under contract to Curtin University. We acknowledge the iVEC Petabyte Data Store, the Initiative in Innovative Computing and the CUDA Center for Excellence sponsored by NVIDIA at Harvard University, and the International Centre for Radio Astronomy Research (ICRAR), a Joint Venture of Curtin University and The University of Western Australia, funded by the Western Australian State government.

\bibliographystyle{mn2e}
\bibliography{mn-jour,mwa}

\label{lastpage}

\end{document}